\documentstyle[preprint,12pt,prd,aps]{revtex}
\draft
\tightenlines
\begin{document}
\title{QUASI-SUPERSYMMETRY IN TOP SECTOR OF THE STANDARD MODEL}
\author{B. B. Deo and L. P. Singh}
\address{Department of Physics \\ Utkal University \\ Vani
Vihar \\ Bhubaneswar 751 004, India}
\maketitle 
%\narrowtext
\begin{abstract}
We show the existence of quasi-supersymmetry as formulated
by Nambu in the top-sector of the standard model. We
present the explicit form of the quasi-supersymmetric
charge. We also deduce,like Nambu, a quasi supersymmetric mass relation
 which is the same as  suggested by Veltman.
\end{abstract}

\newpage
Higgs fields provide the only missing link in the standard
model of Glashow, Weinberg and Salam which has satisfied
the experimental tests to a high degree of accuracy. To
cure the standard model of this deficiency, Nambu [1] has
been pursuing the idea of bootstrap symmetry breaking
mechanism for last few years wherein the Higgs boson itself
is the cause of the attractive interaction between formions
that leads to symmetry breaking and dynamical generation of
Higgs as a Cooper pair bound state of fermions through a
BCS mechanism. In this picture, there are fermionic and
bosonic low-energy excitations. The bosonic excitations
exist in Goldstone $(\pi)$ and Higgs $(\sigma)$ modes. It
is a characterstic of BCS mechanism that the Goldstone,
fermion and Higgs modes satisfy simple mass relations
\[ m_{\pi} : m_{f} : m_{\sigma} = 0 : \Delta : 2 \Delta =
0:1:2 \]
where $\Delta$ is the energy gap parameter.

These low energy modes can be represented by
Ginzburg-Landau-Gell-Mann-Levy Hamiltonian in which the
bososn self-coupling and the boson-fermion Yukawa coupling
are related so as to satisfy the above mass ratio. Nambu
has further shown that the static part of GL Hamiltonian
can be factored as an anticommutator of fermionic
operators as
\begin{equation}
\bar{H}_{st} = [ \bar{Q}, \bar{Q}^{+} ]_{+}
\end{equation}
much like in supersymmetry. (The bar indicates spatial
integral.) Since the $Q s$ and $Q^{+} s$ are not nilpotent,
Nambu cahracterises such theories as quasi-sypersymmtric.
Eventhough, it is not clear why the BCS-mechanism should
have a built-in quasi-supersymmetry, the above
factorisation also works in relativistic field theories.

It is crucial in these constructions, as one learns by
working out the anticommutators explicitly, that the
fermionic and bosonic degrees of freedom must match because
the fermion kinetic energy comes weighted by number of
boson field contractions and boson kinetic energy piece
receives its weightage from the number of fermion field
contractions [2].

We, in what follows, wish to provide an explicit example of
the above ideas using the standard electroweak unification
model. We assume, following Nambu [3], that the Higgs boson
interacts strongly with top quark only so that it can be a
composite of top quark (t) and anti-top quark $(\bar{t})$.
The top quark and the Higgs bosons play a key role in the
electroweak unification and their interactions deserve a
separate attention. This should make good sense now that the
top quark is found to be very heavy implying large
Yukawa coupling and hence strong top-top interaction. The
BCS quasi-supersymmetry, as stated earlier, predicts a 2:1
mass ratio between the Higgs and the top quark. The
bootstrap condition is formulated by requiring cancellation
of  quadratic divergence of the tadpole diagrams
representing Higgs induced attractive interaction among
fermions and the contribution of Higgs and gauge boson
loops. The logic for requirement of cancellation of
quadratic divergence being that eventhough the
Salam-Weinberg theory is the low energy effective theory of
a more fundamental theory without Higgs, it should be
independent of the underlying high-energy scale carried by
the quadratic cut-off. The absence of quadratic and
logarithmic divergences leads to two kinds of mass relations
involving the masses of top quark, Higgs and gauge bosons
like [4],
\begin{equation}
12 m_{t}^{2} = 3 m_{H}^{2} + 3 m_{Z}^{2} + 6 m_{W}^{2}
\end{equation}
and
\begin{equation}
12 m_{t}^{4} = \frac{3}{2} m_{H}^{4} + 3 m_{Z}^{4} + 6 m_{W}^{4}
\end{equation}
A more constraining mass relation has also been obtained
by Deo and Maharana [5] as
\begin{equation}
12 m_{t}^{4} = m_{H}^{4} + 3 m_{Z}^{4} + 6 m_{W}^{4}
\end{equation}
requiring the vanishing of quadratic divergence and
cosmological constant upto one-loop. One solution of the
above mass relation gives Higgs mass to be nearly the twice
of top quark mass reminding us of the underlying BCS
quasi-supersymmetry. Such a guess is further supported by
the fact that in a theory involving Higgs fields, gauge
fields and the top sector of the quark fields, there is an
exact match of fermionic and bosonic degrees of freedom as
top quarks have twelve degrees of freedom, four vector
mesons have eight and the a complex Higgs scalar doublet
has four defrees of freedom.

We now proceed to demonstrate factorisability of the top
sector of the standard model Hamiltonian in terms of supersymmetric
charges. 

To set up the notations, we first note that the free
Hamiltoninan density of the t-sector of the standard model
has the following form
\begin{equation}
{\cal H} = {\cal H}_{g} + {\cal H}_{q} + {\cal H}_{s}
\end{equation}
where ${\cal H}_{g}, {\cal H}_{q}$ and ${\cal H}_{s}$ are
the Hamiltonian densities involving the gauge fields,
quarks and scalar fields respectively which in turn have
the following form in temporal gauge $(W_{0}^{i} = 0 \;
{\rm and} \; Z_{0} = 0)$;
\begin{eqnarray}
{\cal H}_{g} &=& \frac{1}{2} \left [ \sum_{i = 1,2,3}
\left ( \vec{E}_{i}^{2} + \vec{B}_{i}^{2} \right ) + \left
( \vec{\cal E}^{2} + \vec{\cal B}^{2} \right ) \right ] \nonumber \\
{\cal H}_{q} &=& i \sum_{a=B,Y,R} \sum_{i=L,R}
 t_{i}^{a+} \vec{\sigma} \cdot \vec{D}_i t_{i}^{a} \nonumber 
\end{eqnarray}
and 
\begin{equation}
{\cal H}_{s} = \pi^{+} \pi + (\vec{D} \Phi)^{+} \cdot (\vec{D}
\Phi) + V_{cl} 
\end{equation}
In Eqn. (6) $\vec{E}_{i}, \vec{B}_{i}$ and $\vec{\cal E}$
and $\vec{\cal B}$ refer to electric and magnetic fields of the
SU(2) gauge bosons $W_{i = 1, 2, 3}$ and $U(1)$ gauge boson
$Z$, $t_{L,R}$ refer to left-handed and right-handed
$t$-quark, $a$ stands for colour indices blue (B), yellow
(Y) and red (R) and finally $\Phi$ stands for a complex
Higgs doublet $(\phi,\varphi )$ and $\pi$ for its conjugate
momenta. The $V_{cl}$ is given by $V_{cl} = \lambda (\mid \Phi
\mid^{2} - \sigma^{2}/2)^{2} \equiv W^{2}$ where $\sigma =
246 GeV$ is the electro-weak symmetry breaking scale.

The covariant derivatives $\vec{D}_{L}$, $\vec{D}_{R}$ and
$\vec{D}$ acting on left handed t-quark,right-handed t-quark and Higgs field
$\Phi$ respectively, have the form,
\begin{eqnarray}
\vec{D}_{L} && = \left ( \vec{\nabla} + \frac{ig^{\prime}}{6} {\vec{Z}} +
\frac{ig}{2} {\vec{W}_{3}} \right ) \nonumber \\
\vec{D}_{R} && = \left ( \vec{\nabla} +\frac{ 2ig^{\prime}}{3}
{\vec{Z}} \right ) \nonumber \\
\vec{D} && = \left ( \vec{\nabla} + \frac{ig^{\prime}}{2} {\vec{Z}} +
\frac{ig}{2} {\tau_{i} \vec{W}_{i}} \right )
\end{eqnarray}
$\sigma_{i}$ and $\tau_{i}$ are Pauli matrices, designated
differently as they operate in spin and weak-isospin space
respectively. 

The supersymmetric charge $Q_{\alpha}$ whose anticommutator
with its hermitian conjugate $Q_{\alpha}^{+}$ (here sum
over index $\alpha$ i.e. trace of anticommutator matrix is
implied) correctly reporduces the Hamiltonian given by
equations (5) and (6) is found to be
\begin{eqnarray}
Q && = (\pi_{\phi} + \vec{\sigma} \cdot (\vec{\nabla} +
\frac{ig}{2} \vec{W}_{3} + \frac{ig^{\prime}}{2} \vec{Z}) \phi^{+}
+ \frac{ig}{2} \vec{\sigma} \cdot
\vec{W}_{-} \varphi^{+}) {t_{L}}^{\prime Y} \nonumber \\
&& + ( \pi_{\varphi} + \vec{\sigma} \cdot (\vec{\nabla} -\frac{ig}{2}
\vec{W}_{3} + \frac{ig^{\prime}}{2} \vec{Z} ) \varphi^{+} +
\frac{ig}{2}\vec{\sigma} \cdot \vec{W}_{+} \phi^{+} ) t_{L}^{\prime B}
\nonumber \\
&& + \vec{\sigma} \cdot \vec{F}_{3+} t_{L}^{\prime R} +
\frac{\vec{\sigma}}{ \sqrt{2}} \cdot (\vec{F}_{1+} + i \vec{F}_{2+})
t_{R}^{\prime Y} +\frac{ \vec{\sigma}}{ \sqrt{2}}  \cdot (\vec{F}_{1+} -
i \vec{F}_{2+}) t_{R}^{\prime B} \nonumber \\
 && + (\vec{\sigma} \cdot \vec{F}_{+} +{\eta} W) t_{R}^{\prime R}
\end{eqnarray}
where $\pi_{\phi}$ and $\pi_{\varphi}$ are momenta conjugate
to fields $\phi$ and $\varphi.~t^{\prime}_{L,R}$ are Wilson
line-transformed left-handed and right-handed t-quark
fields defined as,
\begin{eqnarray}
t_{L}^{\prime}(x) && = \exp [\frac{ ig}{2} \int_{0}^{x} \vec{W}_{3}.
d\vec{Y} + \frac{ ig^{\prime}}{6} \int_{0}^{x} \vec{Z} \cdot d
\vec{Y} ] t_{L} (x) \nonumber \\
{\rm and} \; \; t^{\prime}_{R} (x) && = \exp [ i
\frac{2g^{\prime}}{3} \int_{0}^{x} \vec{Z} \cdot d \vec{Y}
 ] t_{R} (x) \nonumber \\
{\rm and } \; \; \mid \eta \mid && = 1 
\end{eqnarray}
Finally $\vec{F}_{i+} = ( \vec{E}_{i} + i \vec{B}_{i}) / \sqrt{2}$ and $\vec{F}_{+} = ( \vec{\cal{E}} + i
\vec{\cal{B}} )/ \sqrt{2}$ are linear combinitions of
electric and magnetic fields produced by the gauge fields
$\vec{W}_{i}$ and $Z$ respectively. The Wilson lines are
introduced as prescribed by Nambu [6] to recover the
covariant derivatives in the Hamiltonian. Thus
\begin{equation}
[ \bar{Q}_{\alpha}, \bar{Q}_{\alpha}^{+} ]_{+} = 2H
\end{equation}
where $H = \int d^{3} x  {\cal H}$.

Going a step further, we find that for $W = 0$ the
anticommutator of $\bar{Q}_{\alpha}$ with
$\bar{Q}_{\beta}^{+}$ (i.e. the full anticommutator $ 2
\times 2$ matrix not just its trace) yields,
\begin{equation}
[ \bar{Q}_{\alpha}, \bar{Q}_{\beta}^{+} ]_{+} = ( \sigma_{\mu}
P_{\mu} )_{\alpha \beta}
\end{equation}
where $\sigma_{0}$ is just the unit matrix and $P_{0}$ is
the Hamiltonian. $\vec{P}$ denotes the three-momentum
vector deduced from the Lagrangian which contains only the
top quark, vector bosons and the Higgs doublet.

In arriving at eqn. (11) we have used the well known trick
of writing [7]
\begin{equation}
(D\phi)_{i}^{+} (D\phi)_{j} = \lim_{i \rightarrow j} \vec{\nabla}_{i}
\vec{\nabla}_{j} S(x,y)
\end{equation}
where the nonlocal string operator $S(x,y)$ has the form
\begin{equation}
S(x,y) = \phi^{+}(y) \exp [ i \alpha \int_{x}^{y} \vec{A}
\cdot d \vec{\zeta} ] \phi(x).
\end{equation}
where $\vec{A}$ stands for a generic gauge field. Thus the
cross term of the form $\vec{\sigma} \cdot
(\vec{D}\phi)^{+} \times (\vec{D}\phi)$ obtained in
computation of the anticommutator is written as
$\epsilon_{ijk} \sigma_{i} (\vec{D} \phi)_{j}
(\vec{D}\phi)_{k}$ can be seen to vanish with the use of
eqn. (12) and eqn. (11) results.

Finally we wish to derive the mass sum rule (Eqn. 2) by
adding the Yukawa term to scalar part of the Hamiltonian
such that the potential in the top sector in the broken
$SU(2) \times U(1)$ phase with $\Phi = ( 0, 0, 0, \chi/
\sqrt{2})$ ($\chi$ real) becomes
\begin{eqnarray}
V && = \frac{\lambda}{4} (\chi^{2} - \sigma^{2})^{2} +
\frac{1}{4} g^{2} W_{\mu+}  W_{\mu-} \chi^{2} \nonumber \\
&& + \frac{(g^{2} + g^{\prime 2})}{8} Z_{\mu}^{2} \chi^{2}
+ g_{y} (t_{L}^{+} t_{R} + t_{R}^{+} t_{L}) \chi
\end{eqnarray}
It needs to be pointed out here that the Yukawa term in
Eq. (14) is quasi-SUSY invariant. This invariance is easily
verified by using the standard variation formula for
fermion fields $\Psi (t_{L}, t_{R})$ and scalar field $\chi$
as \\

$ \delta \Psi_{L}^{+} = [ \bar{\epsilon} \cdot Q,
\Psi_{L}^{+}] = \bar{\epsilon}_{\alpha} [
Q_{\alpha}, \Psi_{L}^{+} ]_{+}  $~~
and
~~$ \delta{\chi} = [ \bar{\epsilon} \cdot Q, \chi ]
= \bar{\epsilon}_{\alpha}  [ Q_{\alpha}, \chi ]$
 \\

with $Q$ given by Eq. (8). The two dimensional equivalence
of $\gamma_{0} \vec{\gamma} \Psi_{L,R}$ with $\pm
\vec{\sigma} \Psi_{L, R}$ and the mass term $\bar{\Psi} \Psi
\longrightarrow \Psi_{L}^{+} \Psi_{R}$ $+ \Psi_{R}^{+}
\Psi_{L}$  in two component notation is used.
The mass operator squares for Higgs and Goldstone bosons
are, as usual, given by
\begin{eqnarray}
m_{H}^{2} && = \frac{\partial^{2} V}{\partial \chi^{2}}
= 3 \lambda \chi^{2} - \lambda \sigma^{2} \nonumber \\
m_{G}^{2} &&  = \lambda (\chi^{2} - \sigma^{2})
\end{eqnarray}
leading to sum of mass operator squares of spin-0 particles
(one Higgs and three Goldstone bosons)
\begin{equation}
\sum m_{0}^{2} = 6 \lambda \chi^{2} - 4 \lambda \sigma^{2}
\end{equation}
Similarly sum of mass operator squares of spin-1 particles
(2W's and 1Z) and of spin-1/2 particles (3 coloured t-quark
and 3 coloured $\bar{t}$-quark) can be obtained as
\begin{equation}
\sum m_{1}^{2} = g^{2}\chi^{2} /2 + (g^{2} + g^{\prime
2})^{1/2}\chi^{2} /4
\end{equation}
and 
\begin{equation}
\sum m_{1/2}^{2} = 6 g_{y}^{2} \chi^{2}
\end{equation}
In order to eliminate the various coupling constants we
scale the fields like
\begin{eqnarray}
\chi && = \chi^{\prime}/(2 \lambda)^{1/2}, W_{1,2} = (2
\lambda)^{1/4} W_{1,2}^{\prime}/g \nonumber \\
Z && = Z^{\prime} (2\lambda)^{1/4} /(g^{2} + g^{\prime
2})^{1/2}, t_{L, R} = (2 \lambda)^{1/8}
t_{L,R}^{\prime} /g^{1/2}_{y} 
\end{eqnarray}
and obtain the scaled potential as
\begin{equation}
V^{\prime} = \frac{1}{8} (\chi^{\prime 2} - (2 \lambda)^{1/2}
\sigma^{2})^{2} + \chi^{2} W_{\mu - }^{\prime} W_{\mu
+}^{\prime} + \frac{1}{2} Z_{\mu}^{\prime 2} \chi^{\prime
2} + (t_{L}^{\prime +} t_{R}^{\prime} + t_{R}^{\prime +}
t_{L}^{\prime}) \chi^{\prime}
\end{equation}
Now, we obtain, in the usual notation,
\begin{eqnarray}
\sum m_{0}^{\prime 2} && = 3 \chi^{\prime 2} -
2(2\lambda)^{1/2} \sigma^{2} \nonumber \\
\sum m_{1}^{\prime 2} && = 2 \chi^{\prime 2} (W) +
\chi^{\prime 2} (Z) \nonumber \\
\sum m_{1/2}^{\prime 2} && = 6 \chi^{\prime 2} \nonumber
\end{eqnarray}
leading to
\begin{equation}
\frac{\partial}{\partial \chi^{\prime 2}} ( \sum
m_{0}^{\prime 2} + 3 \sum m_{1}^{\prime 2} - 2 \sum
m_{1/2}^{\prime 2} ) = 0
\end{equation}
It is interesting to note that this sum rule is reminiscent
of the supersymmetric one obtained by Ferrara, Giradello
and Palumbo [8] like
\begin{equation}
\sum_{J} (-1)^{2J} (2J +1) M_{J}^{2} = 0
\end{equation}
Noting the invariance of traces of the mass operator square
sum-rule for both primed and unprimed states which are
scaled with respect to each other by constant factors, we
obtain, 
\begin{equation}
\frac{\partial}{\partial \chi^{2}} \left [ \sum m_{0}^{2}
(\chi) + 3 \sum m_{1}^{2} (\chi) - 2 \sum m_{1/2}^{2}
(\chi) \right ] = 0
\end{equation}
The above relation implies the following relation between
the coupling constants :
\begin{equation}
6 \lambda + 3(g^{2}/2 + (g^{2} + g^{\prime 2})^{1/2} /4) -
12 g_{y}^{2} = 0
\end{equation}
Multiplying $\sigma^{2}$ to the above relation and using
\\

$ m_{H}^{2} = 2 \lambda \sigma^{2}, m_{W}^{2} = g^{2}
\sigma^{2}/4, m_{Z}^{2} = (g^{2} + g^{\prime 2})^{1/2}
\sigma^{2}  /4$ and $m_{t}^{2} = g_{y}^{2} \sigma^{2}$ 
\\

one immediately
obtains the required mass sum rule of Veltman [9], as a
consequence of factorisability of classical potential as in
case of supersymmetric theory,
\begin{equation}
12 m_{t}^{2} = 3 m_{H}^{2} + 3 m_{Z}^{2} + 6 m_{W}^{2}
\end{equation}
For large Higgs mass, $m_{H} \simeq 2 m_{t}$ typical of a
$\langle \bar{t} t \rangle$ condensate. 

Our derivation of Veltman sum rule follows the elegant
method of Einhorn and Jones [10].
There is, by assumption, no $\langle \bar{b} b \rangle$
condensate. The Higgs boson and the breaking of the $SU(2)
\times U(1)$ symmetry is due to $\langle \bar{t} t \rangle$
consensate [11]. Therefore, it is the top sector which has
been considered in this letter.

Eqn. (11) being in the form of exact supersymmetry algebra,
signifies existence of an underlying sypersymmetry in the
top-sector of the standard Weinberg-Salam model. This
symmetry, as we showed, also provides the underlying reason
for the Veltman-like mass square sum rule in the sector
under consideration. In the absence of any experimental
confirmation of existence of supersymmetric particles,
existence of such a deeper symmetry as explicitly exhibited
by us, principally originating from matching of fermionic
and bosonic degrees of freedom in certain sector of a model
deserves serious attention as a viable alternative picture.


\begin{references}
\bibitem{} Y. Nambu, EFI preprint 90-37, Okubo Festshrift
(1990). 
\bibitem{} Y. Nambu, Supersymmetry and Quasi-supersymmetry,
Essay in honour of M. Gell-Mann (1991).
\bibitem{} Y. Nambu, EFI preprint 90-46, R. Dalitz
Festschrift (1990); EFI preprint 89-08 (unpublished).
\bibitem{} See for example Ref. 1.
\bibitem{} B. B. Deo and L. Maharana, Phys. Lett. {\bf
B323}, (1994) 417.
\bibitem{} Y. Nambu, BCS mechanism, Quasi-supersymmetry and
fermion masses (unpublished).
\bibitem{} See for example H. Bohr and G. Rajasekaran,
Phys. Rev. {\bf D32}, (1985), 1547.
\bibitem{} S. Ferrara, L. Giradello and F. Palumbo, Phys.
Rev, {\bf D20}, (1979), 403.
\bibitem{} M. Veltman, Acta Phys. Polon, {\bf B12}, (1984),
437.
\bibitem{} M. B. Einhorn and D. R. T. Jones, Phys. Rev. D {\bf
46},(1992) 5206.
\bibitem{} See for example J. D. wells, Phys. Rev. D {\bf
56}, (1996), 1504; W. A. Bardeen, C. T. Hill and M.
Lindner, Phys. Rev., {\bf D 41}, (1990), 2647.
\end{references}
\end{document}